\documentclass{PoS}

\title{A Very Forward Hadron Spectrometer for the LHC and Cosmic Ray Physics}

\ShortTitle{Very Forward Hadron Spectrometer}

\author{\speaker{Michael Albrow}\thanks{Scientist emeritus}\\
        Fermi National Accelerator Laboratory, Batavia, IL 60510, USA.\\
	ORCID 0000-0001-7329-4925 \\
        E-mail: \email{albrow@fnal.gov}}

\abstract{Charged hadron production in hadron-hadron collisions with longitudinal momentum fraction
Feynman-x, $x_F$, between 0.1 and 0.9  has not been measured above $\sqrt{s}$ = 63 GeV at the CERN
Intersecting Storage Rings. I discuss a way to measure this at the Large Hadron Collider at
$\sqrt{s}$ = 13 TeV, which is 40,000 times higher in equivalent fixed target energy, and important for understanding
cosmic ray showers. }

\FullConference{2nd World Summit: Exploring the Dark Side of the Universe\\
		25-29 June, 2018\\
		University of Antilles, Pointe-à-Pitre, Guadeloupe, France}

\begin{document}

\section{Introduction}

When the first hadron collider, the CERN Intersecting Storage Rings, ISR, came into operation in 1971 
it was known that most produced particles in inelastic collisions have small transverse momenta, 
$p_T = \sqrt{p_x^2 + p_y^2}$ less than about 1 GeV/c. Feynman had proposed, based on the idea of 
parton constituents of protons, that their longitudinal momenta, $p_z$, should scale with the beam momentum, 
so the spectra at high energies should be a function only of $x_F$ and $p_T$ where Feynman-x, $x_F = p_z/p_{BEAM} = 2 \times p_z/\sqrt{s}$.
 With $p_{BEAM}$ ranging from 11.5 GeV/c to 31.4 GeV/c at the ISR most produced particles would be at very small angles. 
 The Small Angle Spectrometer, Experiment R201, was designed to measure the spectra of identified 
 $\pi^+,\pi^-,K^+,K^-,p, $and $\bar{p}$ in this region, with $p_T <$ 2.5 GeV/c \cite{sasisr1, sasisr2}.
 There are no measurements at higher $\sqrt{s}$, which is 200 times higher at the LHC than at the ISR, with the 
 exception of Roman pot devices measuring diffractively scattered protons with $x_F >$ 0.9, and 
 neutral particles, mainly neutrons and $\pi^0 \rightarrow \gamma\gamma$, in calorimeters at $\theta$ = 0$^{\circ}$. The  
 $\pi^+$ and $\pi^-$ spectra are bound to be very different from each other (with two u-quarks and only one 
 d-quark in the proton) with $\pi^0$ in-between. Similarly, we expect $K^+ > K^-$ and $p > n > \bar{p}$, etc. 
 We should measure all of these charged particles at the LHC, and can do so with high precision with new detectors.
 Decaying particles such as $K^0_s \rightarrow \pi^+\pi^-$ and $\Lambda \rightarrow p \pi^+$ may be accepted, but especially
 interesting would be $D^0 \rightarrow K^{\pm} \pi^{\mp}$ and $J/\psi \rightarrow \mu^+\mu^-$.
 
  \begin{figure}[t]
  \vspace{-1.4 in}
 \centering
\includegraphics[angle=0,origin=c,width=120mm]{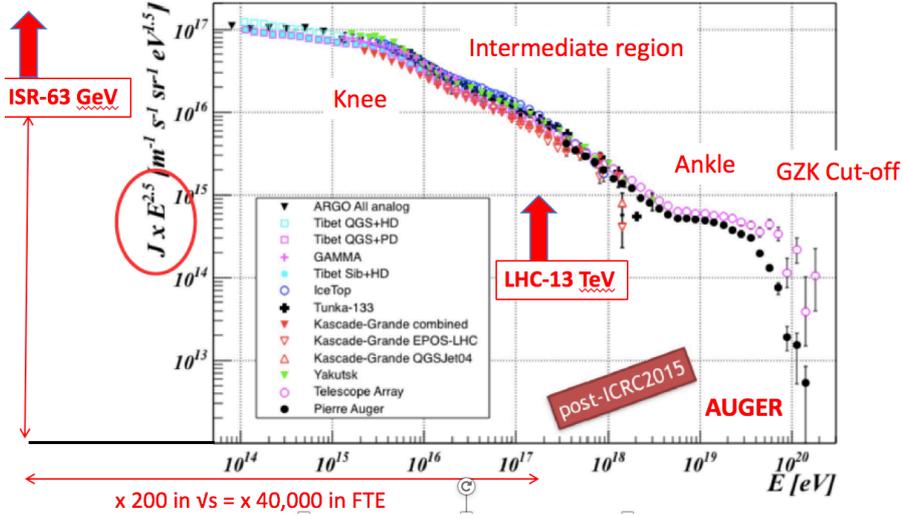}
 \vspace{-1.75 in}  

\caption{The flux of cosmic rays multiplied by $E^{2.5}$, showing the ``knee" and the ``ankle" vs. primary
energy $E$. All this data is based on measuring showers in the atmosphere. The maximum energy of the CERN ISR and LHC are
indicated.}
 \end{figure}

The flux of hadrons or nuclei in cosmic rays at such high energies is $ <$ 1 $m^{-2}$ per year, 
so direct detection, which would have to be from space, is impractical. Therefore, all our knowledge of cosmic ray spectra,
see Fig. 1,  
from below the ``knee" at $10^{15}$ - $10^{16}$ eV   to the highest energies at the GZK cut-off around $10^{20}$ eV, 
comes from 
showers in the atmosphere. The measurements are calorimetric, e.g. detecting the scintillation 
light (fluorescence) in the atmosphere (a homogeneous calorimeter) or sampling the shower with a ground array of 
scintillation counters or water tanks (which is a sampling calorimeter with a single sparsely-covered layer).
Figure 1 shows the flux of cosmic ray showers, multiplied by $E^{2.5}$ from the top ISR energy, far below the ``knee", and
well beyond the
 LHC energy. There have been speculations that the knee is caused by a change in the nature of the interactions, even perhaps in the
 forward direction, the subject of this note. 
 
Understanding the calibration of these ``calorimetric" detectors, i.e. the relation between measured signal and the energy of the 
incident cosmic ray, is crucial.  The shower process is complicated, from the first interaction on a nucleus of 
nitrogen or oxygen, at energies up to a thousand times higher than those accessible at the LHC, with secondaries, 
tertiaries, etc. down to low energies in a cascade. There are many Monte Carlo programs modelling these inelastic 
interactions, such as {\sc kaskade, dpmjet, epos, qgsjet, pythia} and {\sc sybil}. When these are 
used to predict the spectra of high $x_F$ neutrons and $\pi^0$ at LHC energies they differ by more than an order 
of magnitude. Those are measured \cite{zerodeg} with small calorimeters and are not very precise, but they
serve to show that very forward hadron production is poorly understood. Of course, these simulations assume 
nothing unexpected occurs in small angle hadron production over many orders-of-magnitude in energy. Precise 
measurements, of order 2\% in both momentum and cross-section, of charged hadron spectra are possible, and will enable these various 
models to be rejected, modified or tuned, or they may reveal unexpected phenomena. Importantly, there are some puzzles 
in cosmic ray showers such as an anomalous high content of muons \cite{muonxs}. Muons can come from decays of charm and beauty hadrons which have never 
been measured at high $x_F$, as well as Drell-Yan annihilation of very low-$x_{Bjorken}$ antiquarks with quarks.

     \begin{figure}[!ht]
     \vspace{-1.4 in}
\begin{center}
 \includegraphics[width=0.90\textwidth]{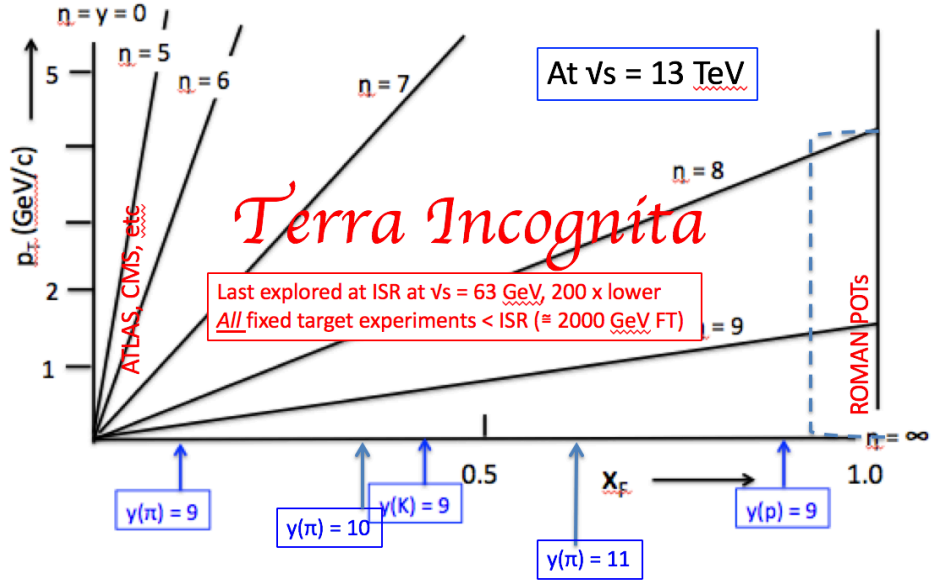}
  
 \vspace{-1.85 in}  
\caption{For $pp$ collisions at $\sqrt{s}$ = 13 TeV, regions of low transverse momentum $p_T$ and all Feynman-x, $x_F$,
showing lines of constant pseudorapidity $\eta$. Protons with $x_F >$ 0.9 are measured in Roman pots, and neutral particles
in calorimeters around $0^\circ$. Identified charged hadrons have not been measured except at $\eta <$ 4 at LHCb, so most of this phase
space is Terra Incognita.}

\end{center} 
 \end{figure}

  Figure 2 shows the region of small $p_T$ and the full range of $x_F$, most of which is \emph{``Terra Incognita"}, not measured above
  $\sqrt{s} = 63$ GeV, except for nearly elastic protons $x_F >$ 0.9 and neutral hadrons in calorimeters.
   
   The most conventional mechanism for producing a high-$x_F$ low-$p_T$ hadron at the LHC is as a product of diffractive 
excitation. The simplest, lowest-mass examples are 
$p^* \rightarrow n \pi^+$ or $p^* \rightarrow p \pi^+ \pi^-$, where $p^*$ is a diffractively excited proton. Such 
processes were measured at the ISR in the Split Field Magnet and, at lower $\sqrt{s}$, in the Omega Spectrometer at the SPS
(fixed target). 
At the LHC the diffractively scattered proton will have $x_F = 1 - M^2(p^*)/s \approx 1.0$ and could be 
measured in a Roman pot, only possible with special high-$\beta^*$ running. There will be a very large 
rapidity gap, with no hadrons, between that scattered proton and the lowest-rapidity hadron from the $p^*$ fragmentation.

There are many reasons, apart from measurements of diffraction fragments, to add a Very Forward Hadron Spectrometer, VFHS, to an (existing) collision region at the LHC. 
It can cover a large region of parameter space ($p_T, x_F, \sqrt{s}$) which is completely unexplored. 
These are ``legacy measurements" and it would be a serious omission to close the LHC without having measured them, not only in $pp$ but in $pA$ and $AA$
collisions.  
They probe non-perturbative QCD, in a region not yet well understood, in novel ways. They will greatly improve our understanding 
of cosmic rays (and hence of the high-energy cosmos) in terms of their spectra and composition, including muon content, e.g. from
heavy flavours.. 
And novel and unexpected phenomena may be revealed.

\section{A possible spectrometer design}

At $\sqrt{s}$ = 13 TeV a hadron with $x_F >$  0.1 and $p_T$ = 1 GeV/c has polar angle $\theta$ < 1/650 = 1.5 mrad. It passes 
down the vacuum pipe, through quadrupole fields and then the (MBX) ``beam separation dipoles", spectrometer magnets with an 
integral field of $B.dL$ = 30 Tesla-meters. Charged hadrons are bent out of the beam. Behind these dipoles is a field-free 
region from about 80 to 140 m, where the TAN absorber, which protects the superconducting LHC magnets, is located. At Point 5 (where 
CMS is located) this is a simple cylindrical pipe, which transitions to two separate pipes for the incoming and outgoing beams 
at 140 m. (Zero-degree calorimeters for neutral particle measurement are located between these pipes.) This pipe can be 
redesigned to allow the deflected hadrons to emerge from the vacuum through little material. The pipe for the outgoing protons may start 
at a $z$-position about 12-15m in front of the TAN, making that much space available for the detectors. The pipe cross section should be as small as 
allowable by the machine operations to maximize the acceptance, with positive hadrons emerging on one side and negative on the other.

   I present the concept with approximate numbers, to show that such a spectrometer is feasible. Cross sections are relatively
   high and running with modest luminosity (pile-up of a few interactions per bunch crossing) is preferred.
   The size 
   of the detectors is small, only tens of cm$^2$ per layer. Immediately outside the vacuum chamber window, precision tracking 
   with silicon strips or pixels measures the straight tracks, to a precision 
   in $x,y,\theta_x, \theta_y$ of about 10 $\mu$m and a few $\mu$r. With the assumption that the track originated in a 
   beam-beam collision, the particle's charge is known and its momentum with a resolution of about 2\%. Non-primary tracks 
   coming from interactions in the beam pipes and other material will generally have forbidden values of 
   $x,y,\theta_x, \theta_y$, and also can mostly be removed by comparing momentum with the energy measured in a calorimeter at 
   the back of the spectrometer.

Between the tracker and the calorimeter one can budget about 8 m of longitudinal space for hadron identification, which 
is the main technical challenge of the detectors, discussed later. The calorimeter, about 2.5 m long and of order 1 m$^2$ 
in area, should fill the transverse area as closely as possible around the outgoing beam pipe. Since it is small compared 
to most LHC calorimeters it can be of high performance, with longitudinal segmentation (e.g. 2 EM segments 
plus 3 HADronic segments) and fine transverse granularity, as in an ``imaging" calorimeter. This should allow a good 
measurement of more than one particle, correction for lateral leakage of showers close to the edge of the fiducial region, 
and distinction between photons, electrons, hadrons and muons. 

Behind the calorimeter, tracking chambers identify muons and signal any hadron shower leakage. Note that 
muons in this 1 - 6 TeV range have radiative energy losses several times higher than minimum-ionizing particles; they are 
on the ``relativistic rise". Good muon measurement is very important, since they can come from heavy flavours $c$ and $b$; the 
branching fraction of $D^0 \rightarrow \mu + X$ is 6.7\%, and leptons also come from $J/\psi$ and $\Upsilon$ decays. Background 
muons from $\pi$- and $K$-decay will be well known since their spectra will be measured, and $\gamma c \tau (\pi)$ = 340 km 
at 2.5 TeV (18.55 km for $K^\pm$)!

Between the upstream tracking and the calorimeter about 8 m of space can be dedicated to charged hadron identification. 
Cherenkov counters cannot separate such high momentum $\pi$ and $K$, but transition radiation detectors which measure 
$\gamma = E/m$ are the most likely (perhaps the only) techniques. Their use in high energy experiments is usually to 
aid in $e/\pi$ separation at low momenta, in a limited space and difficult environment. Motivated by this {\sc vfhs} initiative, R \& D is being 
carried out at CERN \cite{romaniouk} to identify these multi-TeV hadrons. X-radiation from the interface between two materials with 
different dielectric properties has a $\gamma$-dependent probability of order 1/137, and an emission angle $\theta_X$ of 
order 1/$\gamma$. A typical detector has hundreds of radiator foils separated by gaps, and design parameters (foil material 
and thickness, spacing and the gas in the gaps) can be tuned for different $\gamma$-ranges. We need to cover the range from 
the slowest particles accepted (1 TeV protons, $\gamma = E/m $ = 10$^3$) to the fastest (5 TeV $\pi$, $\gamma = 3.6 \times 10^4$). 
A 10 - 20\% $\gamma$ measurement should give adequate $\pi/K/p$ separation. The ``angle-TRD" concept has many short stacks followed by pixel X-ray 
detectors, whose superimposed signals form a ring with radius $\theta_X$. (Unlike Cherenkov radiation, the emission angle 
\emph{decreases} as the particle speed increases.) Together with the X-ray energy measurement, such a detector can minimize the 
material budget for optimum particle identification, and provide the tracking information in addition, alleviating the need for 
a separate tracker.

The detector elements of the VFHS are accessible in the LHC tunnel and modular, so detectors can be replaced as techniques improve. We do not have multi-TeV test beams, so electrons 
and pions are used for tests, which have been done at the CERN SPS \cite{romaniouk, romaniouk2} and compares well with simulations.

Without having a detailed technical design for the spectrometer it appears feasible, with well-understood state-of-the-art 
detectors together with an innovative TRD system that is being developed with very encouraging results. Beam-line and detector 
(\textsc{geant}) simulations, and a realistic vacuum chamber design, are needed for a technical proposal, including 
acceptances for physics processes.

Where could such a spectrometer be located? Of the four existing collision regions, the strongest cases can be 
made to add the VFHS to ALICE or LHCb (or even both!). Relatively low luminosity running is an advantage, as the cross sections are not small. This would give ALICE, whose focus is on heavy-ion collisions, a 
programme of $p+p$ physics orthogonal (in more than one sense!) to the high-$p_T$ central physics done well by ATLAS and CMS. 
The LHC is capable of making both p-N and N-N collisions (after a short machine study), and the VLHC can then measure the
spectra of very forward showers produced in the atmosphere by cosmic rays. The physics focus of LHCb is heavy flavors. Potentially
the VLHC can measure very forward charm and beauty, from ``direct" leptons or possibly $D^0 \rightarrow K^{\pm}\pi^{\mp}$, but acceptance
calculations need to be done for a realistic detector design. Studies of the central event correlated with leading hadrons may
also be interesting.

I apologise for the very incomplete list of citations. I thank Anatoli Romaniouk and his collaborators developing the TRD systems for
VFHS (also known as Small Angle Spectrometer, SAS).
I would like to thank Pierre Petroff and the organisers of ``Exploring the Dark Side of the Universe" for the invitation.

\end{document}